# A Framework for Autonomous Robot Deployment with Perfect Demand Satisfaction using Virtual Forces

Gamal Sallam
Department of Computer and Information Sciences,
Temple University, Philadelphia, PA, USA
tug43066@temple.edu

Uthman Baroudi
Computer Engineering Department
King Fahd University of Petroleum & Minerals
Dhahran, 31261, Saudi Arabia
ubaroudi@kfupm.edu.sa

***Abstract*— In many applications, robots' autonomous deployment is preferable and sometimes it is the only affordable solution. To address this issue, virtual force (VF) is one of the prominent approaches to performing multirobot deployment autonomously. However, most of the existing VF-based approaches consider only a uniform deployment to maximize the covered area while ignoring the criticality of specific locations during the deployment process. To overcome these limitations, we present a framework for autonomously deploy robots or vehicles using virtual force. The framework is composed of two stages. In the first stage, a two-hop Cooperative Virtual Force based Robots Deployment (Two-hop COVER) is employed where a cooperative relation between robots and neighboring landmarks is established to satisfy mission requirements. The second stage complements the first stage and ensures perfect demand satisfaction by utilizing the Trace Fingerprint technique which collected traces while each robot traversing the deployment area. Finally, a fairness-aware version of Two-hop COVER is presented to consider scenarios where the mission requirements are greater than the available resources (i.e. robots). We evaluate our framework via extensive simulations. The results demonstrate outstanding performance compared to contemporary approaches in terms of total travelled distance, total exchanged messages, total deployment time, and Jain's fairness index.**

TABLE I. LIST OF SYMBOLS

| | |
|---|---|
| R, $R_f$ , $R_a$ | Robots, Free robot, Associated robots |
| L | Landmarks |
| $D(L_j)$ | Demand of landmark j |
| $N_r$ ($R_i$) | Neighbor robots of robot $R_i$ |
| $N_l$ ($R_i$) | Neighbor landmarks of robot $R_i$ |
| $w_a$ | Attractive force |
| $w_r$ | Repulsive force. |
| $d_{ij}$ | Distance between robot $R_i$ and robot $R_j$ |
| $d_{th}$ | Distance threshold between robots |
| $\Theta_{ij}$ | Angle between robot $R_i$ and robot $R_j$ |
| $c_{th}$ | Maximum communication range |
| $F_{ij}$ | Force applied on robot $R_i$ from robot $R_j$ |
| $F_{ir}$ | Repulsive force applied on robot $R_i$ from a landmark |
| $F_i$ | The total force applied on robot $R_i$ |

## I. INTRODUCTION

Employing a networked set of robots is an effective way to serve applications in areas where human intervention is impossible or risky. In rescue operations, for example, robots can be used to help in discovering bodies under rubble, and they can even assist the injured. Collaboration among robots will be essential in these applications in order to efficiently achieve the allotted goals in a timely manner. Realizing such collaborative operation without central coordination is a key challenge. Past studies have proposed methods for the distribution of robots, but these have tended to suffer from limitations such as evenly spreading the robots regardless of demand, requiring an a priori demand distribution over an area, or requiring central coordination of the robots.

Blanket coverage refers to the spreading of robots over an area. A number of algorithms have been proposed to achieve maximal blanket coverage based on self-spreading [1–4]. Focused coverage is another model that considers the point of interest (PoI) [5–7, 25]. Assuming a circular (disc) field of view, nodes should be distributed around the PoI to achieve coverage, i.e., the union of the field of view for all nodes is hole-free.

One of the most popular techniques to enable the self-spreading of robots after an ad hoc random placement in an area is to model them as electromagnetic particles that exert virtual forces, which repel or attract neighbors based on proximity [1] [2]. Based on the composite force applied by its neighbors, a robot moves to a new location. This process is repeated many times until the network reaches an equilibrium in which robots become uniformly distributed in the area. This has the following advantages: a simple communication model (size and type of packets), enhancement of the initial coverage degree, control of the coverage degree via the threshold value, fast convergence, and consideration of obstacles, borders, and coverage holes. Other approaches lie in one of the following classes: computational geometry-based, fuzzy-based, and metaheuristic-based [8]. In the computational geometry-based approaches [4,9,10], a geometric computation is used to identify areas with less coverage and to direct the movement of robots from more densely covered areas toward less densely covered areas. Voronoi diagrams and Delaunay triangulation are two common approaches in this class. Weaknesses of this type of approach include the facts that the algorithms are greedy and that they are ineffective when dealing with large holes [11]. In the fuzzy-based approach [12, 13], a fuzzy logic system is used to control robot movement. The fuzzy system applies several rules based on, for example, the Euclidean distance or the number of robots. Then, the system provides a new position to which each robot should relocate. It does not take into account the presence of obstacles. Algorithms belonging to metaheuristic-based approaches utilize the effectiveness of metaheuristics in order to settle the position, direction, and movement speed of a mobile sensor. Ant Colony (AC) [14] and Genetic Algorithms (GA) [15] are examples of such algorithms. These algorithms have high complexity, and the quality of the obtained solutions depends on a large number of parameters (e.g., the number of iterations and GA-related parameters) [8].

In this work, we aim to address the challenges of distributed deployment in scenarios for which a set of pre-installed devices (referred to as landmarks), that are capable to collect some information from their vicinity and make a decision of how much resources are need (demand for robots) within their range, are present in the deployment area. We formulate this dynamic coverage problem using Potential Fields where landmarks and mobile robots exert virtual forces based on the landmarks' demand and the mutual distance between them. The work in [22] addresses this problem, however, it assumes that there is a central gravity point and any landmark with unsatisfied demand will contact this point to get help such that it satisfies its demand. It also assumes that all landmarks are connected. In a recent work [23], all the previous assumptions are removed by the proposed COVER technique, but in this work, we aim to improve COVER in a variety of ways. First, we utilize two-hop communication to reduce the deployment time and travelled distance. Moreover, we introduce a Trace Fingerprint technique that can guarantee the maximum possible demand satisfaction. Finally, we consider the fairness as a selection criterion when distributing robots among landmarks in case the collective demand of all landmarks is greater than the supply of available robots.

## II. LITERATURE REVIEW

The virtual force algorithm (VFA) has been used widely to achieve uniform distributions of robots. In Reference [1], the idea of a virtual force was used for the first time to improve the coverage after a random deployment of mobile sensors. The authors considered a binary detection model in which a target is detected (not detected) with complete certainty by the sensor if a target is inside (outside) its circle. After the initial random deployment, all sensor nodes are able to communicate with the cluster head. The cluster head is responsible for executing the virtual force algorithm and managing the one-time movement of sensors to the desired locations. This work considers a uniform distribution of the mobile sensor. It is centralized in terms of the virtual force calculation, which is a single point of failure. Tan et al. [2] developed a connectivity-preserved virtual force technique such that the covered area is maximized and the connectivity is guaranteed. The developed technique considers that there is a base-station located near the area of interest and the disconnected nodes move toward it to connect.

Wand et al. [16] added a particle swarm optimization (PSO) to the virtual force approach. In the process of self-organized deployment, the nodes do not really move, but the cluster-head node calculates the virtual path first and then guides cluster-in nodes to migrate once to save energy. The fitness function of the PSO is designed to consider the time cost by self-organized deployment and the coverage rate after deployment. Only a uniform distribution of the mobile sensors is considered in this work, and no guidelines are provided for choosing the virtual force parameters. To limit the number of neighboring robots involved in virtual force computation, the authors in Reference [17] suggest the use of Delaunay triangulation. Each robot will only be affected by the attractive force and repulsive force of the nodes that are directly connected to it in the constructed Delaunay triangulation. This approach requires a large amount of computation: each node is required to build a Delaunay diagram for every iteration of the virtual force computation. Garetto et al. [18] proposed a distributed sensor relocation scheme based on virtual forces, adding the restriction that there are at most only six nodes that can exert forces on the current node. This work handled the problem that arises when nodes have a high communication range by restricting the number of nodes to six.

Ying et al. used virtual force for post-deployment to improve the coverage in a wireless sensor network. They assumed that static sensor nodes are initially deployed in the monitoring environment randomly, and the nodes communicate with each other to detect the coverage holes [19]. The mobile nodes will be used to increase the coverage. Assuming that coverage holes generate an attractive field on mobile nodes, the mobile nodes compute the virtual force for many rounds until there is no force toward the mobile node or the maximum number of rounds is reached. The mobile nodes stay where they stop at the last round. If a force is exerted toward a mobile robot from multiple directions, it will cause the robots to oscillate and trigger many unnecessary movements. The same steps are performed in Reference [20] with the mobile robots also using particle swarm optimization to reposition themselves to best cover a sensing hole. Reference [21] considers both obstacles and preferential areas. The obstacles exert a repulsive force based on a rank given to each obstacle, while the preferential areas and target points exert an attractive force based also on a rank given to each preferential point. This work depends on a cluster head to perform all related calculations needed for robot deployment.

In our work, we modify the virtual force such that it accounts for the criticality of each preferential point (landmark) based on cooperation between landmarks and robots. The cooperation is based on the number of landmarks, their demands and the local demand in the range of each robot and landmark. Thus, this work is an improvement on the work in [23]: it overcomes some limitations, such as the expected deadlock when VFs are equal and the robot under consideration will stay in its position, and to improve the performance of the VF, especially in terms of demand satisfaction, total time, and total travelled distance.

## III. TWO-HOP COVER

### A. *Problem statement and System-level Assumptions*

We consider an area of interest *A* that has a set of landmarks *L*. The landmarks are used to guide the robot deployment process. A set of landmarks *L'* is equipped with special capabilities, e.g., sensing and computational resources, to enable them to assess the situation in their vicinity and request the presence of a number of robots *(D)* to perform certain tasks. A set of robots *R* is initially randomly deployed in *A*. The goal is to develop a distributed mechanism for robot self-deployment such that the requirements of each landmark are met. The following enumerates key system model assumptions:

1. Each landmark node knows its location.
2. Robots are homogeneous; i.e., they have the same speed, service capabilities, energy supply, etc.
3. Each landmark can request a number of robots depending on the service needs in its area.
4. Landmarks can communicate with each other and exchange information.
5. Each robot knows its initial position.

*6.* The positions of landmarks and their demands are unknown to the robots.

## B. System model

Let $R$ be a set of robots initially dropped at any point in the area of interest. $N$ is the total number of robots. Let $i$ denote each specific robot, where $i = 1, ..., N$. Each robot has a communication range $c_{th}$ within which it can communicate with other robots and landmarks. Let $L$ be a set of landmarks distributed randomly in the area of interest. The number of landmarks is $M$. Let $j$ denote each landmark, where $j = 1, ..., M$. Each landmark has a demand $D(L_j) \geq 0$, and the demand is represented by a number of robots that should be around a given landmark for a given scenario. Any robot can be in one of two states: free or associated. Free robots are those that are not yet associated with any landmark. Let $R_f$ be the set of free robots, initially $R_f = R$. Associated robots are those that successfully became associated with a landmark $(R_i, L_j)$. Let $R_a$ be the set of associated robots. The aim is to make the number of associated robots, $|R_a|$, equal to the total demand of the landmarks, $|R_a| = \sum D(L_j)$. We use $d_{ij}$ to denote the Euclidean distance between node $i$ and node $j$. Let us denote the landmark associated with a robot by $L_{R_i}$ . Each robot $R_i$ has a set of neighbor robots, $N_r(R_i) = \{r_n : d_{in} < c_{th}$, where $n \neq i$, n = 1, ..., N$\}$ and neighbor landmarks $N_l(R_i) = \{l_j : d_{ij} < c_{th}$, j = 1, ..., M, $\}$. The neighboring robots of robot $R_i$ can be either free i.e., a subset of $R_f$, or associated, a subset of $R_a$. The neighbor landmarks can be either satisfied (i.e., $D(N_{L_j}(R_i)) = 0$) or not satisfied (i.e., $D(N_{L_j}(R_i)) > 0$).

## C. Procedure

$R$ robots will be initially dropped at any point in the area of interest. Robots will utilize the virtual force among themselves to spread over the area and to improve the chances of locating landmarks that have demand. Each robot computes the composite virtual force and moves accordingly. In each move, each robot stops for a period to collect messages from other robots and landmarks in order to decide its next step. Each unassociated (free) robot will behave as in algorithm 1. Basically, it will receive two kinds of messages:

1) Messages from other robots that are not associated with any landmarks. These messages are treated normally as in the basic virtual force (i.e., the robots will utilize them to compute either an attractive force or a repulsive force, depending on the distance to the source robot as in equation 1).

2) Messages from landmarks $L_{replies} \subseteq$ L or other robots $R_a^{replies} \subseteq R_a$ that are already associated with a landmark. For each landmark $L_j \in L_{Replies}$, if $D(L_j) > 0$, it will be a demand message of landmark $L_j$. If $D(L_j) = 0$ and $D(N_l(L_j)) > 0$, it will be a list of landmarks and their demands. Otherwise, the message will be a repulsive force. These details are presented in algorithm 4. For each robot $R_i \in R_a^{replies}$, if $D(L_{R_i}) > 0$, it will be a demand message on behalf of landmark $L_{R_i}$. If $D(L_j \in N_l(R_i)) > 0$, then the robot $R_i$ will reply with a list of landmarks and their demands. Otherwise, it will exert a repulsive force to increase the chances that a robot will move in a direction where it can find landmarks with demand. These details are shown in algorithm 5.

The landmarks announce their demands in terms of a specific number of robots. Robot $R_i$ that hears the demand message will add that landmark to a list dl = $\{(L_k, D_k) | k = 1, ..., N_l$, where $N_l = |N_l(R_i)|\}$ in order to respond to the nearest one based on the Euclidean distance. To avoid more robots than needed moving toward one landmark, an association process is proposed. Each robot first sends an association message to the nearest landmark $L_k \in dl$. Then, if $D(L_k) > 0$, it will reply with a confirmation message. Otherwise, it will send a rejection message. If the robot receives a confirmation message, it will not move immediately toward it. Rather, it will stay in its current position until it either determines that none of its neighbor landmarks has a demand or after multiple iterations (i.e., after waiting for a certain time). The logic behind this is that initially the robots are close to each other, and when one of the robots gets associated, the possibility of being needed by its landmark or other neighbor landmarks is high if this robot stays near the other free robots. If the robot receives a rejection message, it will contact the next landmark in its *dl* if it has already heard from multiple landmarks. If the robot fails to associate with any landmark, it will proceed by computing the composite virtual force and move accordingly.

The total force is calculated as follow:

$$F_i = \sum_{j=1, j\neq i}^{k} F_{ij} + \sum_{r=1, r\neq i}^{L} F_{ir} \qquad (1)$$

, where k is the number of neighboring free robots, and $L$ is the number of neighboring associated robots plus the neighboring satisfied landmarks that don't have any neighboring unsatisfied landmarks.

In computing the composite virtual force, we differentiate between two cases. The first case where neighboring robots are not associated with any landmark, then, the calculation goes as the basic virtual force [1] based on (1).

$$F_{ij} = \begin{cases} w_a \left(d_{ij} - d_{th}\right), \ \theta_{ij} & \text{if } d_{ij} > d_{th} \\ 0 \qquad , & \text{if } d_{ij} = d_{th} \\ w_r \dfrac{1}{d_{ii}}, \theta_{ij} + \pi & \text{if } d_{ij} < d_{th} \end{cases} \qquad (2)$$

$$w_a = \left(\frac{d_{th}}{c_{th}}\right) * (N)^{-\alpha} \qquad (3)$$

$$w_r = N^{\alpha} \qquad (4)$$

,where $w_a, w_r$ are the attractive and repulsive forces, respectively, and $d_{ij}$ is the Euclidean distance between robot $i$ and robot $j$ , $d_{th}$ is the distance threshold that should be maintained between any two robots, and $c_{th}$ is the maximum communication range. The second case is where the received messages come from associated robots or from landmarks, the calculation is as follows. For repulsive messages:

$$F_{ir} = \frac{\alpha w_r}{d_{ir}} \qquad (5)$$

$\alpha$ is an arbitrary but predetermined tuning parameter, e.g., $\alpha$ can take the value 2 (in the presented experiments $\alpha$ =3/2), and the “number of robots” represents a number of mobile robots. Increasing the value of $\alpha$ increases the repulsive force and decreases the attractive force and vice versa.

### *D. Algorithms*

In this part, two algorithms are presented: one for free (unassociated) robots and one for associated robots. Then, a third algorithm is presented for the landmarks operation.

**Algorithm 1: Algorithm for the operation of unassociated robots according to Two-hop COVER.**

Robot $R_i$ sends a neighbors position inquiry message
$R_i$ receives position reply messages from neighboring robots and landmarks:
**for** all replies from landmarks **do**
  **if** $D(L_k) > 0$ **then**
    add $L_k$ to the potential demanding landmarks list *dl*.
  **else if** $D(L_k) == 0$ and the demand of its neighbors demand $D(L_a \in N_l\ (L_k)) > 0$ **then**
    add $L_a$ to *dl*
  **else**
    add the $L_k$ to the repulsive force list $F_r$
  **end if**
**end for**
**for** unassociated robots’ replies $R_k \in N_r\ (R_i)$ **do**
  **if** $d_{ij}$ < *Threshold* **then**
    add $R_k$ to the repulsive force list
  **else**
    add the robot $R_k$ to the attractive force list
  **end if**
**end for**
Process *dl* list according to algorithm 2
**if** Failed to associate **then**
  Compute the composite virtual force (VF)
  Compute the new position and relocate to it
**end if**

**Algorithm 2: Algorithm for the processing of potential landmark list *dl*.**

**Repeat**
Choose the nearest landmark ($L_j \in dl$) and associate to it either directly or through another landmark or robot.
**if** succeed **then**
  Mark this robot as associated (i.e. $R_i \in R_a$)
  Change Robot status to associated
  **if** the association was done directly **then**
    The robot stays in its current position
  **else if** The robot associated through another landmark or robot **then**
    Move until gets in range with its associated landmark
  **end if**
**else**:
  Remove this landmark $L_k$ from the list *dl*
**end if**
**until** *dl* is empty

**Algorithm 3: Algorithm for the operation of associated robots according to Two-hop COVER.**

**if** $D(L_{R_i})$== 0 and the demand of its neighbors is zero|| *current time* − *association time* > *threshold* **then**
  relocate to a position determined by its landmark
**end if**
**if** $R_i$ receives position request message **then**
  **if** $D(L_{R_i})$> 0 **then**
    reply with a demand request on behalf of $L_{R_i}$
  **else if** my neighbor landmarks have a demand **then**
    Reply with a list of the landmarks that have a demand
  **end if**
**else if** receives an association request **then**
  Forward it to the required landmark
**else if** receive an association accept or reject from a landmark **then**
  Forward it to the required robot
**else**
  reply with a repulsive force
**end if**

**Algorithm 4: Algorithm for the operation of the landmark according to COVER**

**for all** landmarks ($L_j \in L$) **do**
  **if** $L_j$ receives a position request message from a robot **then**
    **if** $D(L_j) > 0$ **then**
      Reply with its position and demand.
    **else if** $D(L_j) == 0$ & $D(N_l(L_j)) > 0$ **then**
      Reply with a list of the landmarks that have a demand > 0
      Reply with its position and demand.
    **else**
      work as a repulsive force
    **end if**
  **else if** receives associate message **then**
    **if** $D(L_j) > 0$ **then**
      Reply with accept message, and position to come to it.
    **else if** $D(L_j) == 0$ **then**
      Reply with a reject message.
    **else if** the request to a neighbor landmark **then**
      Forward it to the required landmark
    **end if**
  **else if** receive accept or reject message from a landmark to a robot **then**
    Forward it to the required robot
  **end if**
**end for**

*E. Detailed example*

In the scenario presented in Fig 1, we have 10 landmarks numbered 16 to 25 with corresponding demand vector [0, 4, 0, 0, 1, 0, 2, 3, 3, 2]. We placed 15 robots in the center of the area. They will use the virtual force among them in order to spread throughout the area and search for landmarks. Once a robot receives a message from a landmark, it starts responding to it. If it receives from multiple landmarks, it will respond to them, one by one, based on predefined criteria. Here, we consider the distance between the robot and the landmark to be the decisive factor for joining a certain landmark. In this scenario, all the landmarks will have their demands met. All the landmarks will be able to obtain their demands from the first slot. Only landmarks 17 and 20 will have to wait until robots come into their range. Robots 11-15 will use the virtual force for many slots until robot 14 moves into range of landmark 20 and associates with it. Additionally, robot 15 moves into the range of landmark 17 and associates with it. It also helps other robots (11, 12, and 13) to associate with landmark 17.

## I. TRACE FINGERPRINT TWO-HOP COVER

In the previous section, we introduced Two-hop COVER algorithm, but we still see that even though the level of demand satisfaction is high, it may fail in some scenarios to reach 100% demand satisfaction. In this section, we will try to improve the performance of the Two-hop COVER to guarantee perfect demand satisfaction given that the number of robots is enough to meet the demand of all landmarks. A naive solution is to let the remaining robots wander the whole area randomly until they locate a landmark with demand. However, this approach could take a long time to satisfy the demand, and it may not always reach a 100% demand satisfaction. If the area is large, robots may not travel to areas with demand. Additionally, if the total demand is less than the number of available robots, then there will be no demand for some robots. In that case, the robots will continue to move randomly searching for a demand (which does not exist) until they deplete their energy. These random movements could result in a very high cost in terms of traveled distance and the time needed to locate an unsatisfied landmark. Thus, to reduce such randomness and guide the remaining robots throughout the deployment area until they find an unsatisfied landmark, we propose that each robot visit only the locations that it has not visited before. We do that by keeping a trace of the history of visited places and communicating with other neighboring robots and landmarks to collect more information about their current positions and their traces, if applicable. Then, each robot uses that information to guide its movements while searching for landmarks that are still have unsatisfied demand. We call this approach “Trace Fingerprint”.

After applying Two-hop COVER algorithm, if a robot fails to associate itself with any landmark and the following conditions are satisfied, the robot moves to implement Trace Fingerprint algorithm, which is introduced here. The conditions are: 1) there is no force applied on the robot for many iterations (e.g., 10 iterations, depends on system parameters such as the size of the area and the speed of the robots); and 2) the only force exerted toward the robot is a repulsive force for many iterations (e.g., 15 iterations), following which the robot moves to implement Trace Fingerprint.

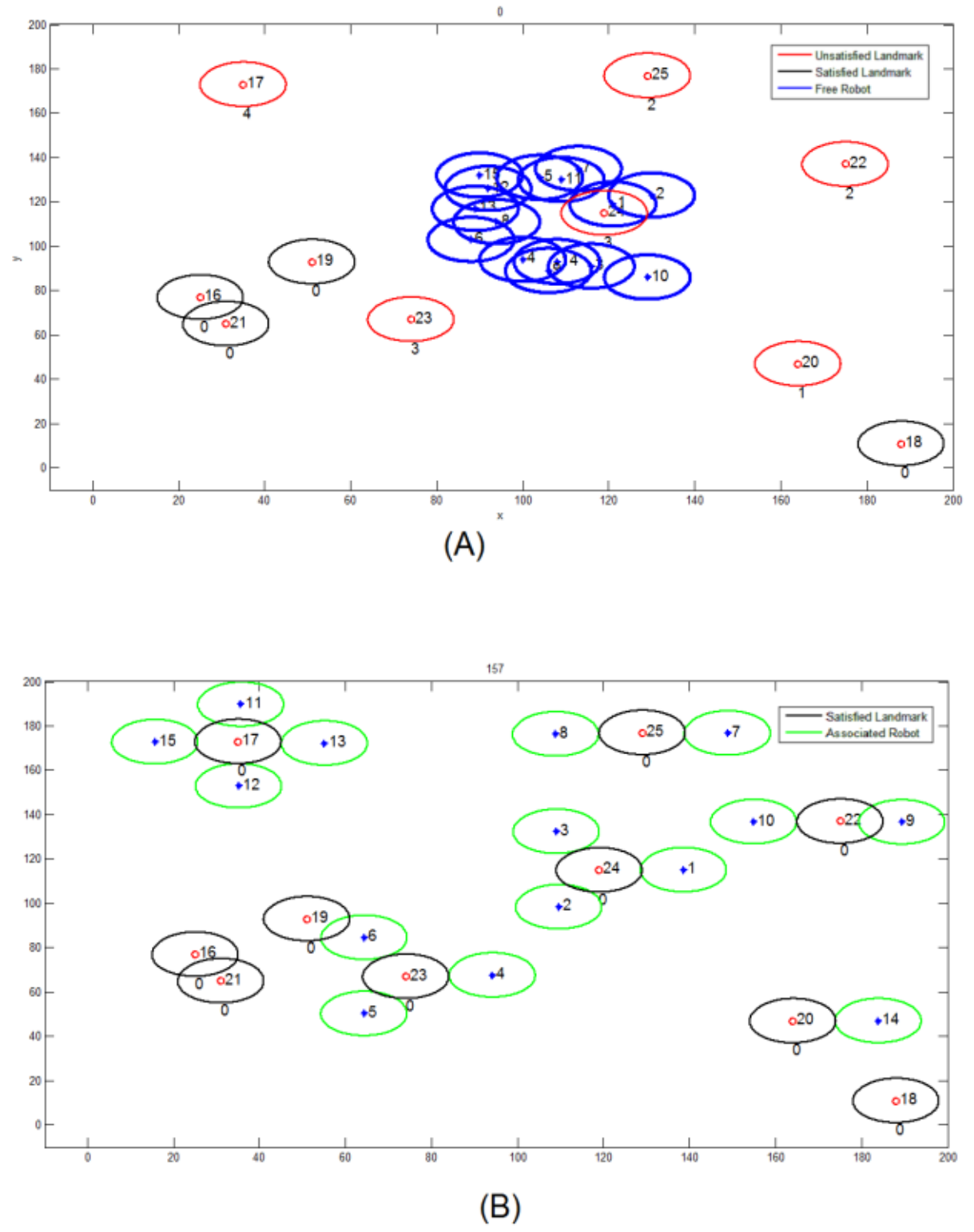

Figure 1. An example of the cooperative landmarks and robots using the virtual force. The number below each red circle is the demand of that landmark. (A) The initial positions of the robots and the demands of the landmarks. (B) The final positions of the robots and the remaining demands of the landmarks.

*A. Detailed example*

Figure 2 shows an example of how to use Trace Fingerprint. In Fig 2-B, according to Two-hop COVER, robot $R_{14}$ is unable to locate any unsatisfied landmarks, so it remains attached to its position with no force applied on it to direct its movement and starts the implementation of Trace Fingerprint. In Fig 2-C, $R_{14}$ divides the area into squares and computes the coverage of each square based on the history it has built. The covered area is shown with a canyon color. Then, the robot moves toward the nearest partially covered square, which is to the right in Fig 2-D. Once the robot reaches its destination, it searches for a landmark, but not finding any; it collects the traces of the new neighbors and repeats the same process. The coverage level computation is repeated again, and the robot finds the nearest uncovered area and moves toward it. The process is repeated until the robot moves into the range of the demanding landmark ($L_{16}$), as in Fig 2-E, and associates with it. Finally, the robot relocates to a position determined by landmark 16, as in Fig 2-F.

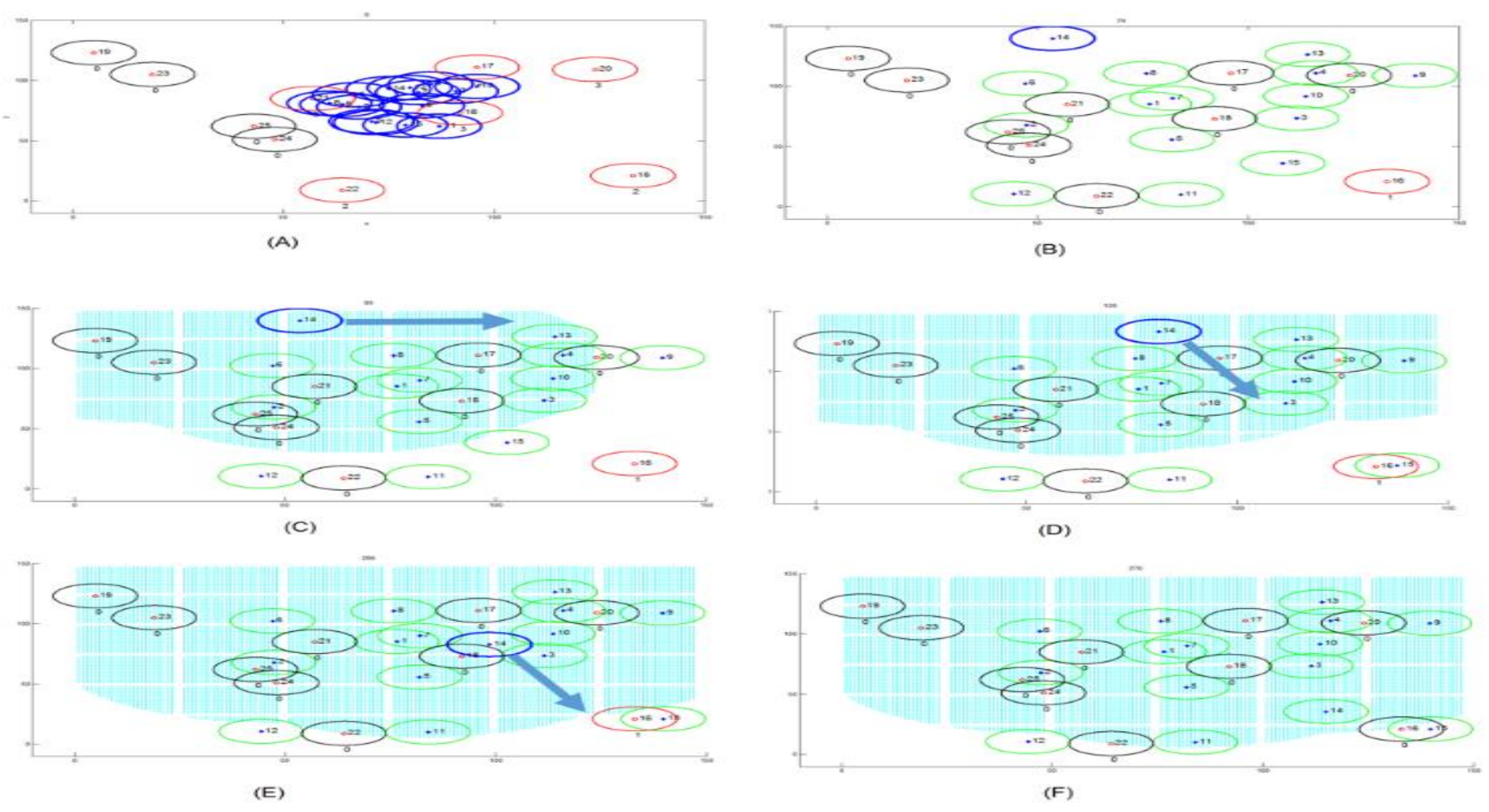

Figure 2 An example of how Trace Fingerprint guarantees 100% demand satisfaction. (A) The initial deployment of the robots. (B) The final deployment according to Two-hop COVER. (C) The start of Trace Fingerprint implementation by robot 14 (Area with a canyon color are covered according to the traces at robot 14). (D) Robot 14 moves toward the nearest free area to the right. (E) Robot 14 moves downward toward the nearest free area and into the range of landmark 16. F) Robot 14 associates with landmark 16.

### *B. Algorithm*

Each robot will maintain a trace set $t_i = \{(x_n, y_n), n = 1...N\}$, where $N$ is the number of points at which the robot stopped and checked for the presence of landmarks, not finding any. Once a robot gets stuck, it will collect the traces of the other robots and build a new one: $T_i = \bigcup_{\forall i}\{t_i\}$, where $i \in \{N_r(R_i)$ and $N_l(R_i)\}$. Each point in the trace will be used to build a virtual map of the places that have been already visited, so $\forall j \in T$, which will be represented in the virtual map as $VM_j$ = a circle centered at $(x_i, y_i)$ with radius $R_c$. Each point within the circle means that there is no landmark on it. The virtual map $VM = \bigcup_{\forall j}\{VM_j\}$, where j $\in T$. Then, the robot will choose the nearest point not covered in *VM* and move toward it. Once it reaches its target, it will check for the presence of a landmark with a demand. If it fails, it will repeat the same process. Algorithm 7 presents the additional steps that will be implemented by a robot when it reaches a stage at which the Two-hop COVER is not helping it to associate with a landmark.

## II. FAIRNESS-AWARE TWO-HOP COVER

Another version of Two-hop COVER is proposed here. It aims to address scenarios in which the collective demand of the landmarks is greater than the available robots. In this case, we do not want the algorithm to be greedy, as in the previous versions. In some applications and real scenarios, it is preferred to ensure that each landmark will get at least portion of its demand, although a high priority can be given to those with high demand . We can achieve this fairness by proposing a minimum demand satisfaction level ($min_{DS}$). When a robot gets close to a landmark that reaches $min_{DS}$, it starts cooperating with its neighboring landmarks so that their level of demand satisfaction is less than the current landmark's demand satisfaction. We measure the level of demand satisfaction by the percentage of the remaining demand over the original demand, i.e., $D_i^{rem}/D_i$, where $D_i^{rem}$ is the remaining demand of robot $i$. In this way, each landmark will secure its minimum level of demand satisfaction and, at the same time, help other landmarks in securing part of their demand. Moreover, when a robot has to decide between which landmark to join, it will use the level of demand satisfaction – not the distance – as a decisive factor to join a certain landmark. To ensure fairness for even the landmarks that are away from the initial positions of the robots, the robots will start virtual force implementation for one iteration without satisfying any landmark's demand. This is solely to increase the chances that each landmark demand will be heard by one of the robots or landmarks and to ensure that each landmark will receive its share of the robots.

**Algorithm 5: Algorithm for operation of Trace Fingerprint**

- For unassociated robots

1: **if** total force ==0 or all force types== repulsive **then**

2: count++

3: **if** count > α **then**

4: change robots status to unassociated searching

5: get the history of neighboring robots and position of landmarks and update history

5: **end if**

6: **else**

7: add my next position to the history

8: **end if**

-For unassociated searching robots

1: **repeat**

divide the area into virtual squares
Compute the coverage level of each square based on the history
make a list of squares that are not fully covered
choose the nearest square and move toward it for a distance *d*
**if** get associated **then**
  change robots status to associated
**else**
  get the positions of the neighbors landmarks and robots and update its history.
**end if**
**until** get associated

## III. Simulation Setup

To evaluate the proposed approaches, we have conducted extensive simulation experiments examining the effectiveness of different setups. The simulation is implemented using Matlab with the parameters in Table II.

The performance metrics used in this study are as follows. 1) Demand satisfaction: this metric measures the percentage of demand that is satisfied by the end of the implementation of the algorithm. 2) The total traveled distance: this metric is used to measure the total distance across which the robots moved in order to achieve the level of demand satisfaction reached by each approach. 3) The total time needed to achieve the demand satisfaction reached by each approach: this is the time until the last associated robot reaches the position determined by its landmark. The time is computed based on a speed of 1 m/s and allowing 3 seconds for each robot to communicate with neighbors at each time a robot finishes one round of movement. 4) Total messages: this metric counts the number of messages that are utilized in the implementation of the Two-hop COVER algorithm. The messages are mainly due to the cooperative virtual force messages.

All above metrics can be used to implicitly measure the energy consumption because the total distance and messages are the main sources of energy consumption.

TABLE II. Simulation Parameters

| **Parameters** | **Value** |
|---|---|
| Simulation tool | Matlab |
| Number of robots (randomly distributed) | 15, 20, 25, 30, 35 |
| Number of landmarks (randomly distributed) | 10 |
| Total landmarks' demand (randomly distributed) | 10, 15, 20, 25, 30, 35 |
| Waiting time | 3 sec |
| Robots' transmission range | 50 m |
| Landmarks' transmission range | 50 m |
| Area size | 150m x 150m, 200m x 200m |
| Stopping criterion | Total force = 0 |

We have compared two-hop COVER with two other approaches namely, the Hungarian algorithm [24] (centralized approach), the COVER [23]. We briefly describe each approach as follows. 1) The Hungarian algorithm (Centralized approach): our problem is that we have a set of resources (robots) and a set of demands of each landmark in terms of robots. The Hungarian method solves this problem by assigning the best robots to each landmark based on the distance between robots and landmarks, and its complexity is $O(n^3)$. 2) COVER: this approach is the one we are making an improvement on it. In COVER, the same procedure of the two-hop COVER is used except that when a robot gets associated (i.e. $R_i \in R_a$) or when a landmark has been satisfied (i.e. $D(L_j) = 0$), it will collaborate with other landmarks that are having demand not satisfied yet (i.e. $d(L_j) > 0$) by applying an attractive force on free robots ($\in R_f$) toward the landmark with the highest demand. Moreover, when a robot gets associated to a landmark, it will immediately move toward it, in opposite to two-hop COVER where the robot will stay in its current position until it either finds out that none of its neighbor landmarks has a demand or after multiple iteration (i.e. wait for a certain time). We compare Trace Finger Print with the Random Waypoint (RWP) approach. Since RWP represents a natural movement for a robot searching for a landmark, we consider RWP as our baseline approach. So, Two-hop COVER will be applied first, if any robot fails to find a landmark with a demand, it will apply Trace Fingerprint or the RWP until it locates a landmark and associate to it.

## IV. Results and Analysis

In order to evaluate the performance of Two-hop COVER and Trace Fingerprint approaches, we have conducted extensive simulation experiments for different scenarios. In both approaches, the number of landmarks randomly distributed over the area is 10. The number of randomly distributed robots varies from 15 to 35. The demand is set to be equal the number of robots.

The results of Two-hop COVER are presented in Figs. 3, 4, 5, and 6. The percentage of demand satisfaction is presented in Fig 3. We see that Two-hop COVER is able to reach a level of demand satisfaction approximately 97% of the level achieved by the centralized approach. This is due to the utilization of the two-hop communication, which allows the robots to reach landmarks that are out of their ranges. In addition, the associated robot stays for a period in its position immediately after becoming associated, which is assumed to help in using the two-hop communication to satisfy the demand of its landmark. We see that Two-hop COVER reaches approximately 100% demand satisfaction, especially when the number of robots is high compared to the area size, and better than in the original COVER.

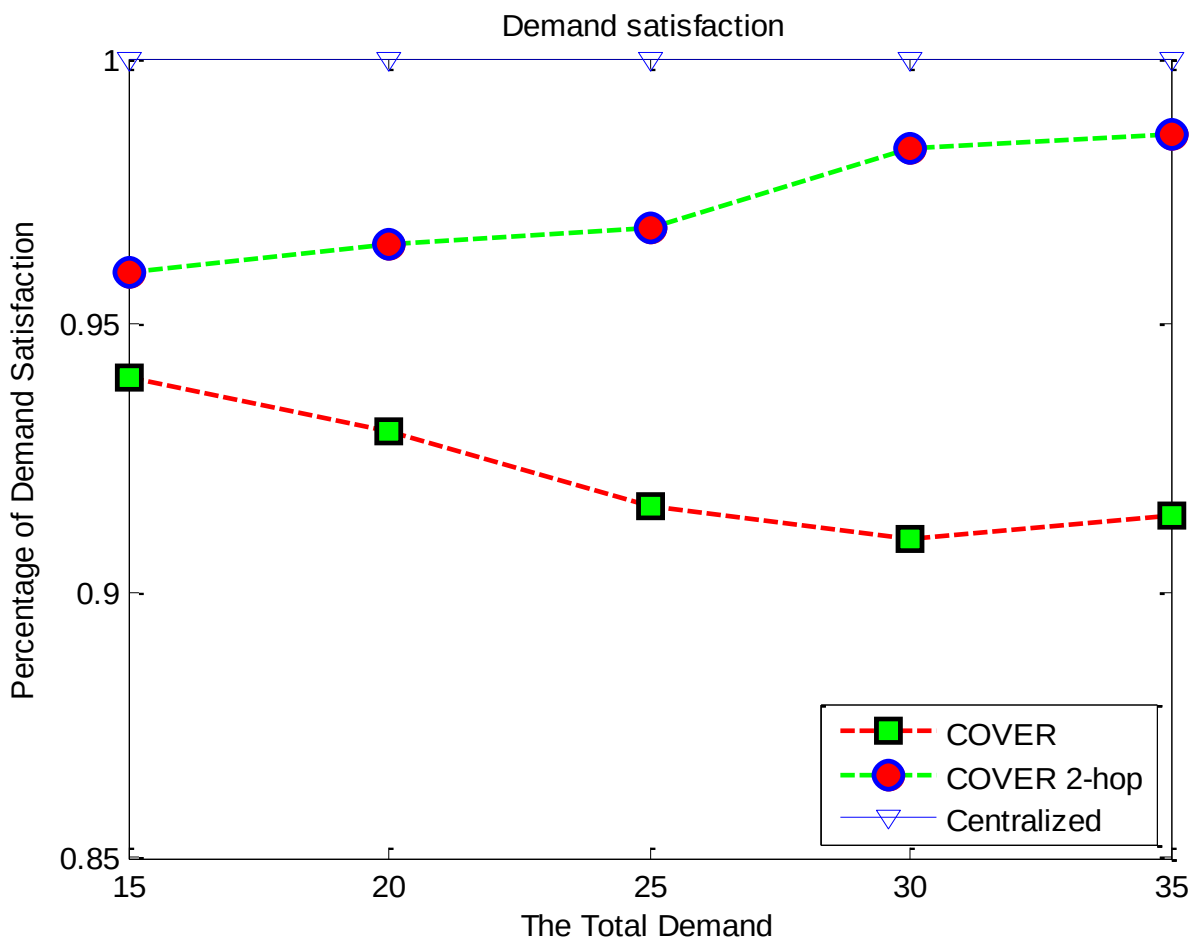

Figure 3. The percentage of demand satisfaction The number of robots equals the total demand, area size=150m x 150m, communication range=50m.

For the total traveled distance, we show a normalized distance which is the total distance compared to the total demand satisfaction, i.e., the average distance travelled for each unit of demand satisfaction. We can see that Two-hop COVER reduced the travelled distance compared to COVER as shown in Fig. 4.

Another factor is presented in Fig 5, which is the total time needed to achieve the level of demand satisfaction in Fig 3 and relocate to the positions determined by the landmarks. Although we introduce a waiting time for the associated robots if their landmark's demand is still not satisfied, Two-hop COVER is able to reduce the total time compared to the original COVER by approximately 20-30%: Two-hop COVER uses two-hop communication, which reduces the main time used to search for a landmark. Additionally, the Two-hop COVER needed an increase of approximately 40% in the total time compared to the centralized approach. The improvements of Two-hop COVER over COVER as percentages are shown in Table III. We can see that Two-hop COVER reduces the total time by approximately 16% when the number of robots is 15 and by 27% when the number of robots is 35. This occurs because increasing the number of robots increases the chances of using two-hop communication between robots and unsatisfied landmarks. Additionally, Two-hop COVER requires an increase in total time of approximately 50% compared to the centralized approach.

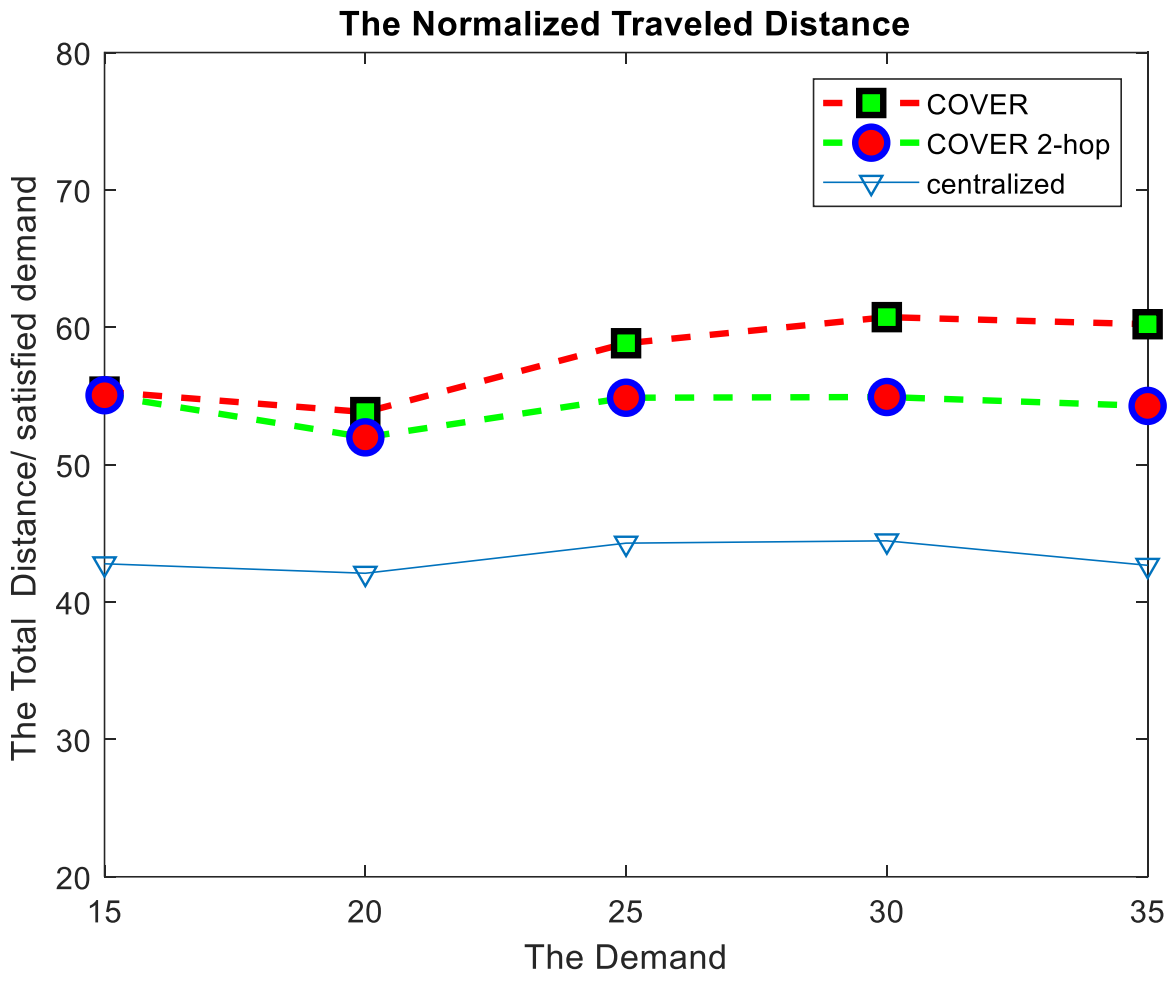

Figure 4. The total traveled distance/demand satisfaction. The number of robots equals the total demand, area size=150m x 150m, communication range=50m.

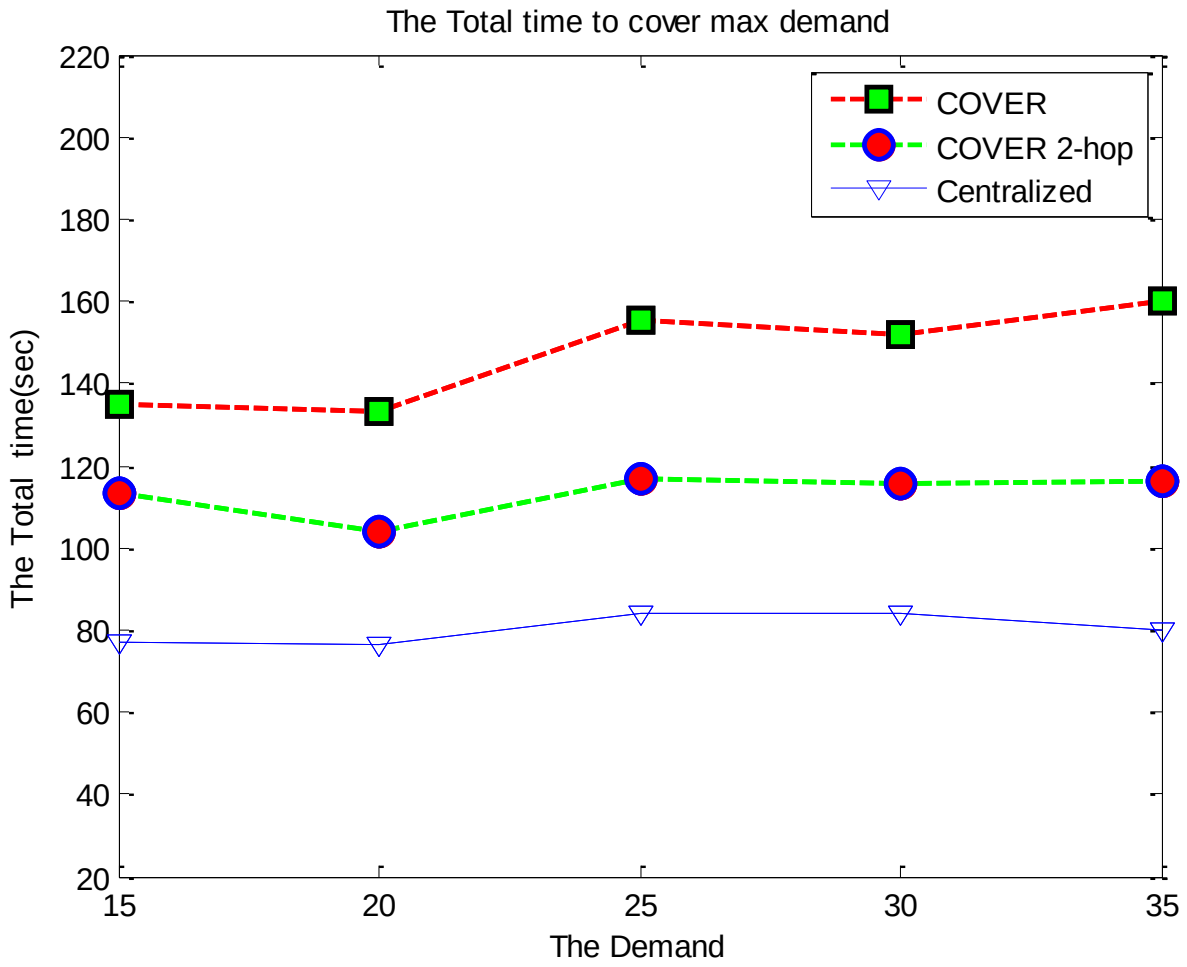

Figure 5. The total time needed to reach the achieved demand satisfaction. The number of robots equals the total demand, area size=150m x 150m, communication range=50m.

TABLE III. THE TOTAL TIME IN SECONDS FOR EACH APPROACH FOR EACH NUMBER OF ROBOTS. THE IMPROVEMENTS OF THE TWO-HOP COVER COMPARED TO THE CENTRALIZED AND COVER METHODS ARE SHOWN AS PERCENTAGES.

| No. of Robots | Centralized | COVER | Two-hop COVER | % of Two-hop VS. Centralized | % of Two-hop VS. COVER |
|---|---|---|---|---|---|
| 15 | 77.20 | 134.80 | 113.50 | +47% | -16% |
| 20 | 76.20 | 133.10 | 104.20 | +37% | -22% |
| 25 | 84.30 | 155.10 | 116.90 | +39% | -25% |
| 30 | 84.00 | 152.10 | 115.50 | +38% | -24% |
| 35 | 80.20 | 160.10 | 116.30 | +45% | -27% |

The total number of messages is the last factor to present in this study. Although Two-hop COVER uses two-hop associations, it succeeded in reducing the total messages exchanged by approximately 40-50% compared to COVER because the faster the robot gets associated, the fewer messages are used. The associated robots will not broadcast any virtual force messages, and consequently, no replies will be needed. In this way, the total messages were considerably reduced as shown in Fig 6.

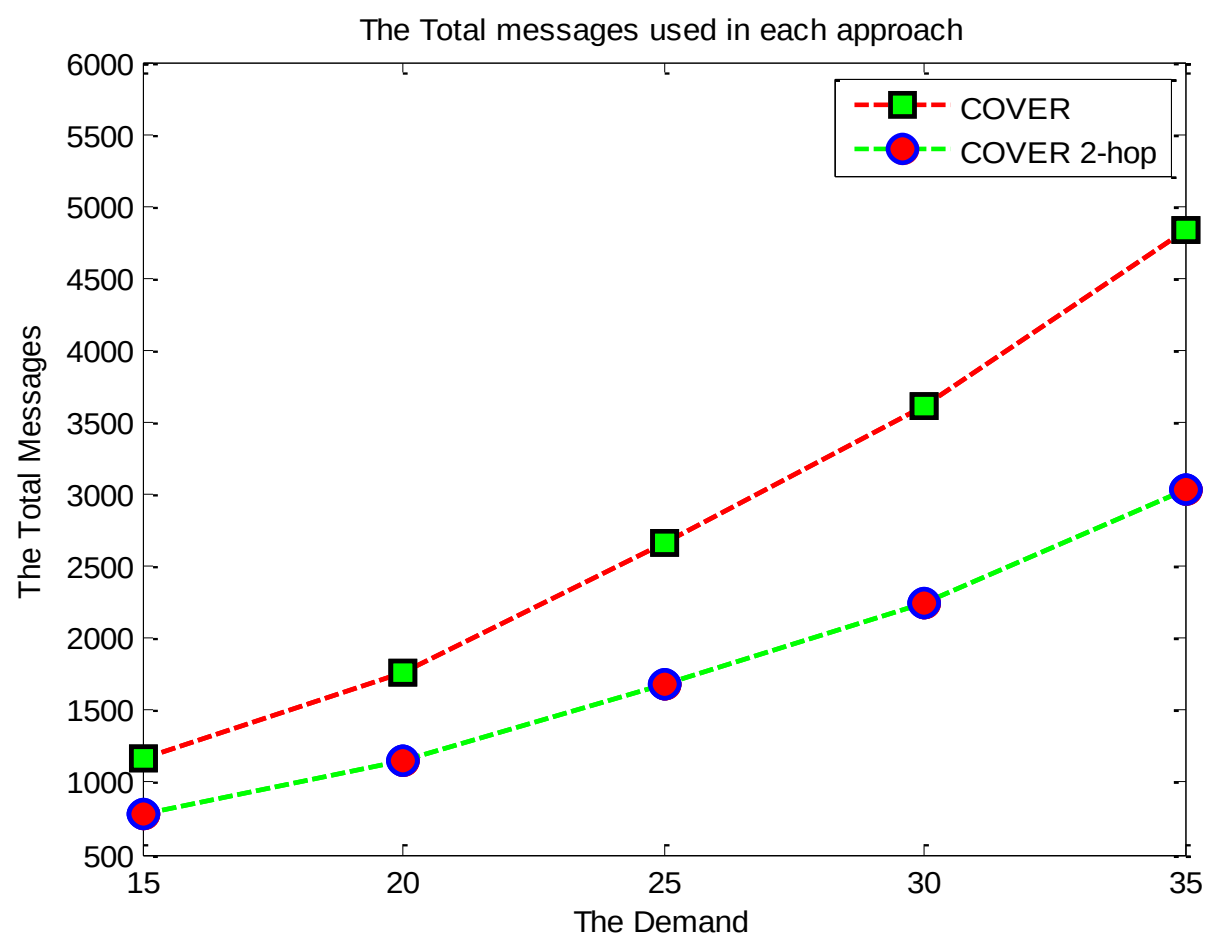

Figure 6. The total number of messages used in each approach. The number of robots equals the total demand, area size=150m x 150m, communication range=50m.

For Trace Fingerprint, we can see in Fig 7 that the Trace Fingerprint succeeded in reaching 100% demand satisfaction. The same level of demand satisfaction was also achieved by the RWP. However, the differences are shown in figure 8, 9, and 10 with respect to the total time, distance, and messages. Trace Fingerprint needed additional movements in order to locate landmarks, but these movements were small compared to those needed for the RWP, as shown in Fig 8. While Trace Fingerprint caused an increase of approximately 20-30% in the distance to reach 100% demand satisfaction compared to Two-hop COVER, RWP caused an increase of more than 100% in the distance traveled. This shows the effectiveness of the proposed approach to guide the robots' movements rather than allowing them to move randomly. In Trace Fingerprint, a robot will not visit previously visited places by it or by its neighbor robot, which increases the chances of locating landmarks quickly and consequently reduces the total traveled distance. Additionally, Table IV shows the percentage of distance needed by the Trace Fingerprint and RWP compared to the centralized approach. We see that Trace Fingerprint needs to travel approximately 70% more distance compared to the centralized method. However, the RWP needs more than 150% when the number of robots is 15 and 100% when the number of robots is 35.

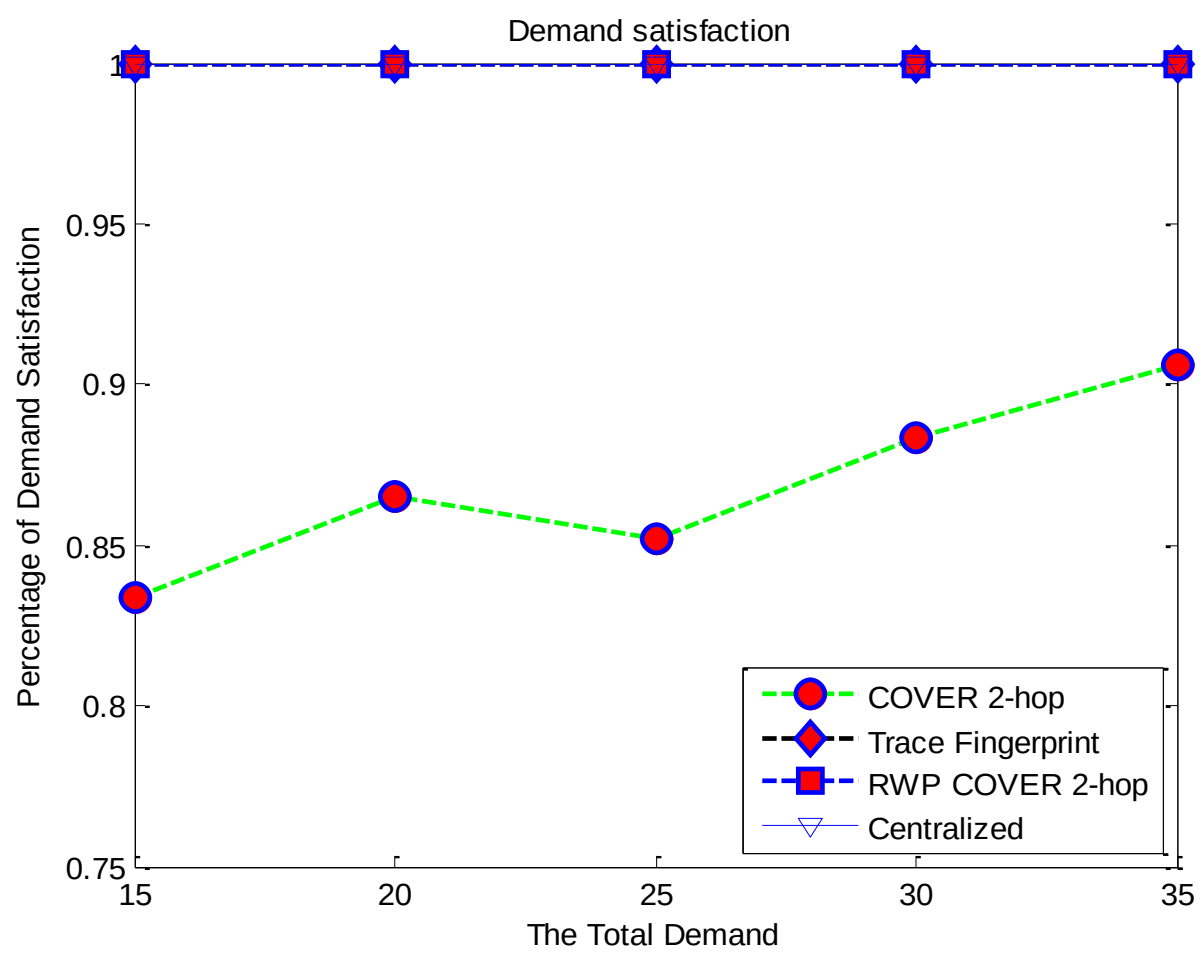

Figure 7. The percentage of demand satisfaction, area size = 200m x 200m, communication range=50m, the number of robots= the demand .

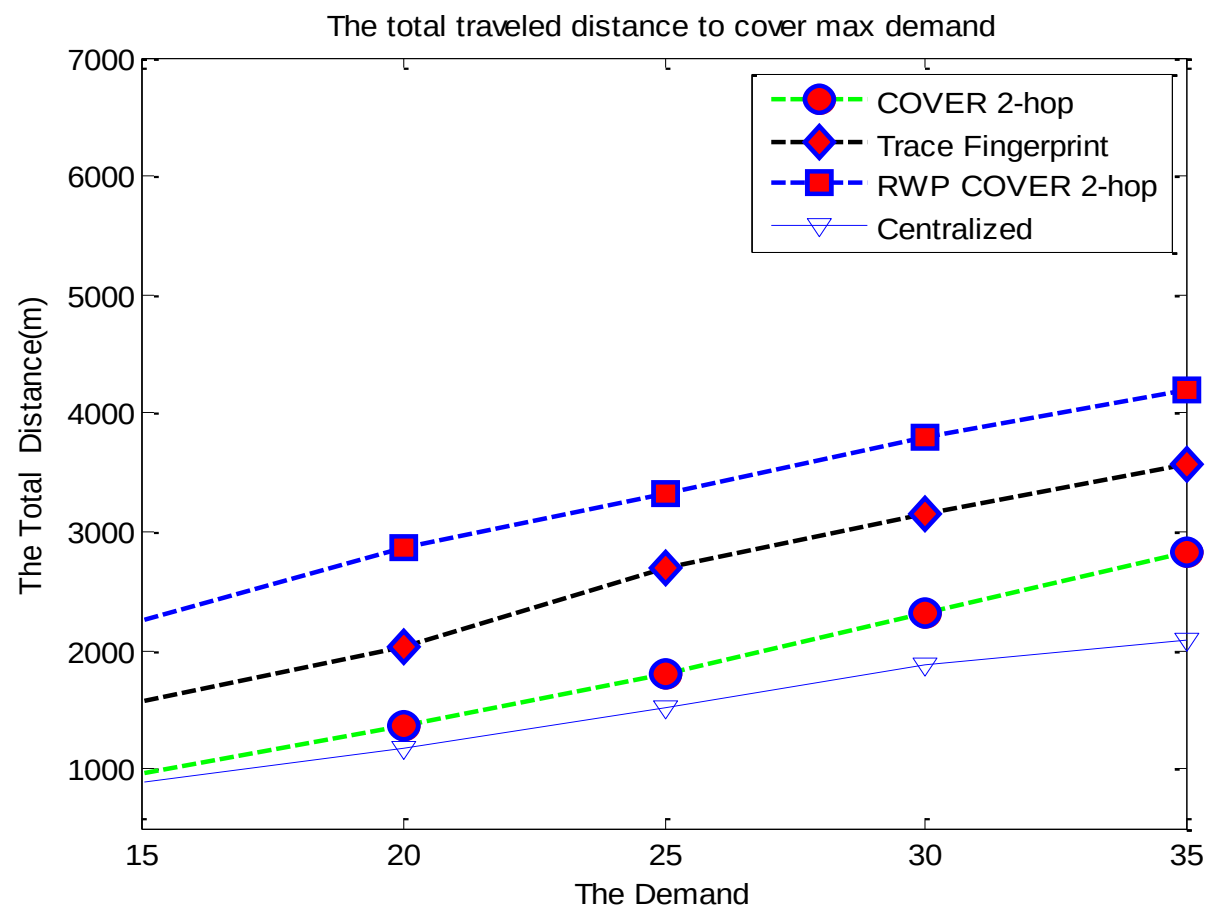

Figure 8. The total distance, area size = 200m x 200m, communication range=50m, the number of robots= the demand.

TABLE IV. THE TOTAL TRAVELED DISTANCE OF TRACE FINGERPRINT COMPARED WITH THE CENTRALIZED AND RWP METHODS. THE PERCENTAGES ARE CALCULATED BASED ON THE DIFFERENCE FROM THE CENTRALIZED APPROACH

| Robots | Centralized | Fingerprint | % | RWP | % |
|---|---|---|---|---|---|
| 15 | 897.60 | 1582.30 | 76% | 2256.70 | 151% |
| 20 | 1181.50 | 2026.30 | 72% | 2871.70 | 143% |
| 25 | 1522.50 | 2701.50 | 77% | 3325.10 | 118% |
| 30 | 1882.50 | 3153.40 | 68% | 3804.10 | 102% |
| 35 | 2083.60 | 3572.30 | 71% | 4199.80 | 102% |

For the total time, due to the guided movements in Trace Fingerprint, the total time needed to reach 100% demand satisfaction is very small compared to that of the RWP, as shown in Fig 9. Finally, the longer it takes to reach the maximum demand satisfaction, the higher the number of messages that will be used. Thus, we see that RWP requires more messages than Trace Fingerprint, as seen in Fig 10. The better performance of Trace Fingerprint compared to RWP comes at a computational cost. Each free robot will need to make a calculation that is not needed in RWP. The robot will need to store each location's history and the locations of its neighbors. Moreover, at each iteration, the robot will need to find the coverage level of each square in order to decide its next move. For time critical applications and with the current advances in computational resources, we can ignore such computational overhead and consider only other factors (time, distance, and messages).

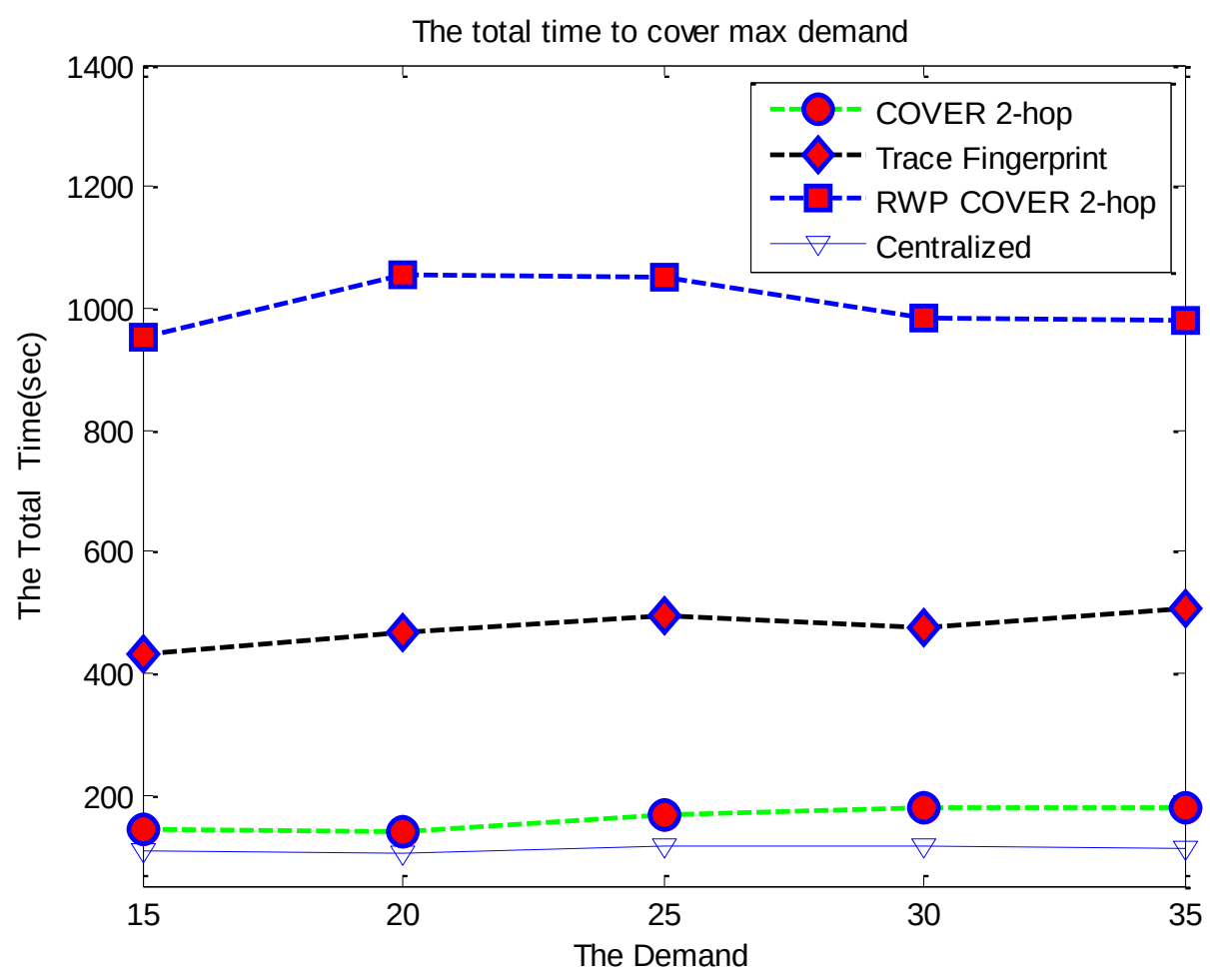

Figure 9. The total time, area size = 200m x 200m, communication range=50m, the number of robots= the demand

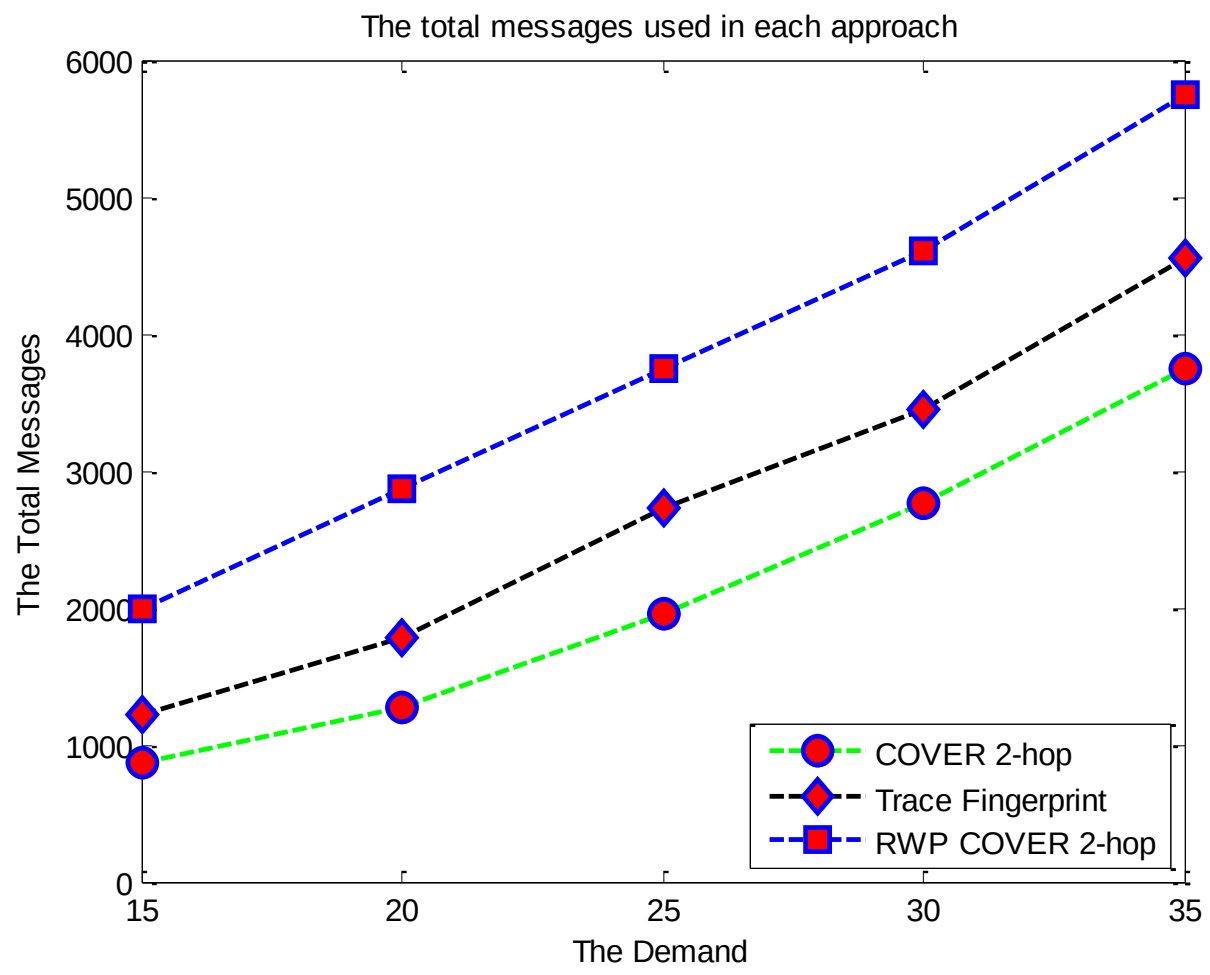

Figure 10. The total number of messages, area size = 200m x 200m, communication range=50m, the number of robots equals the demand

Providing fair dispatching of robots among different landmarks is very important and desirable in practical scenarios. In order to explore the level of achieved fairness, we adopted Jain's fairness index as metric for the percentage of the satisfied demand of the landmarks in all studied approaches, according to Eq. 6.

$$J = \frac{\sum_{i=1}^{N}(D_i - D_i^{rem})^2}{N\,(\sum_{i=1}^{N}(D_i - D_i^{rem}))^2} \quad (6)$$

Where $D_i^{rem}$ is the total remaining demand of robot *i*.

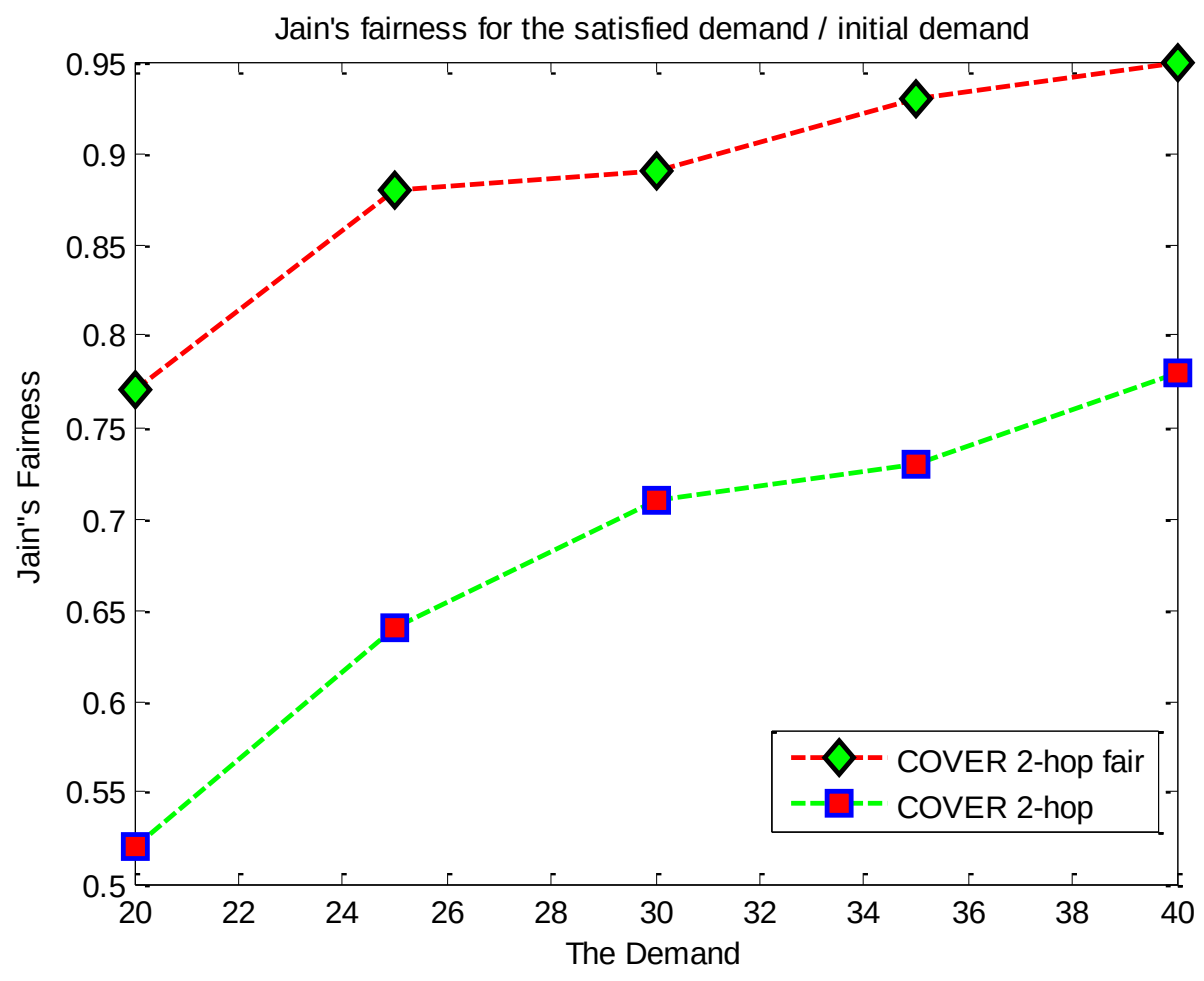

Figure 11: Jain’s fairness index of the percentage of the satisfied demand of the landmarks. The umber of robots = demand=10.

The minimum satisfaction level ($min_{DS}$) is set to be equal to 50% of the original demand. Figure 11 shows Jain’s fairness index. Jain’s fairness in Fairness-aware Two-hop COVER is closer to 1 than that of Two-hop COVER. The fairness increases with the increase in the demand because the increase in the demand implies that the number of robots and the probability of all landmarks securing part of their demand are high. However, the introduction of fairness affected other metrics, such as the total time, distance, and messages. The level of demand satisfaction does not change, as in Fig 12, but the total distance increases in the fairness-aware version by approximately 30%, as shown in Fig 13. The increase in the total distance occurs because each robot considers only the highest demanding landmark with which to associate. Thus, if there are two landmarks, one nearby with a small demand percentage and the other far away with a high demand percentage, the robot will choose the farthest one to associate with, which consequently causes a long distance to be traveled. Moreover, the Fairness-aware approach utilizes virtual force at the beginning to spread the robots over a larger area in order to increase the chances of hearing the demands of all of the landmarks. This leads to an additional distance and a greater amount of time compared with Two-hop COVER. The same holds for the total time, as shown in Fig 14.

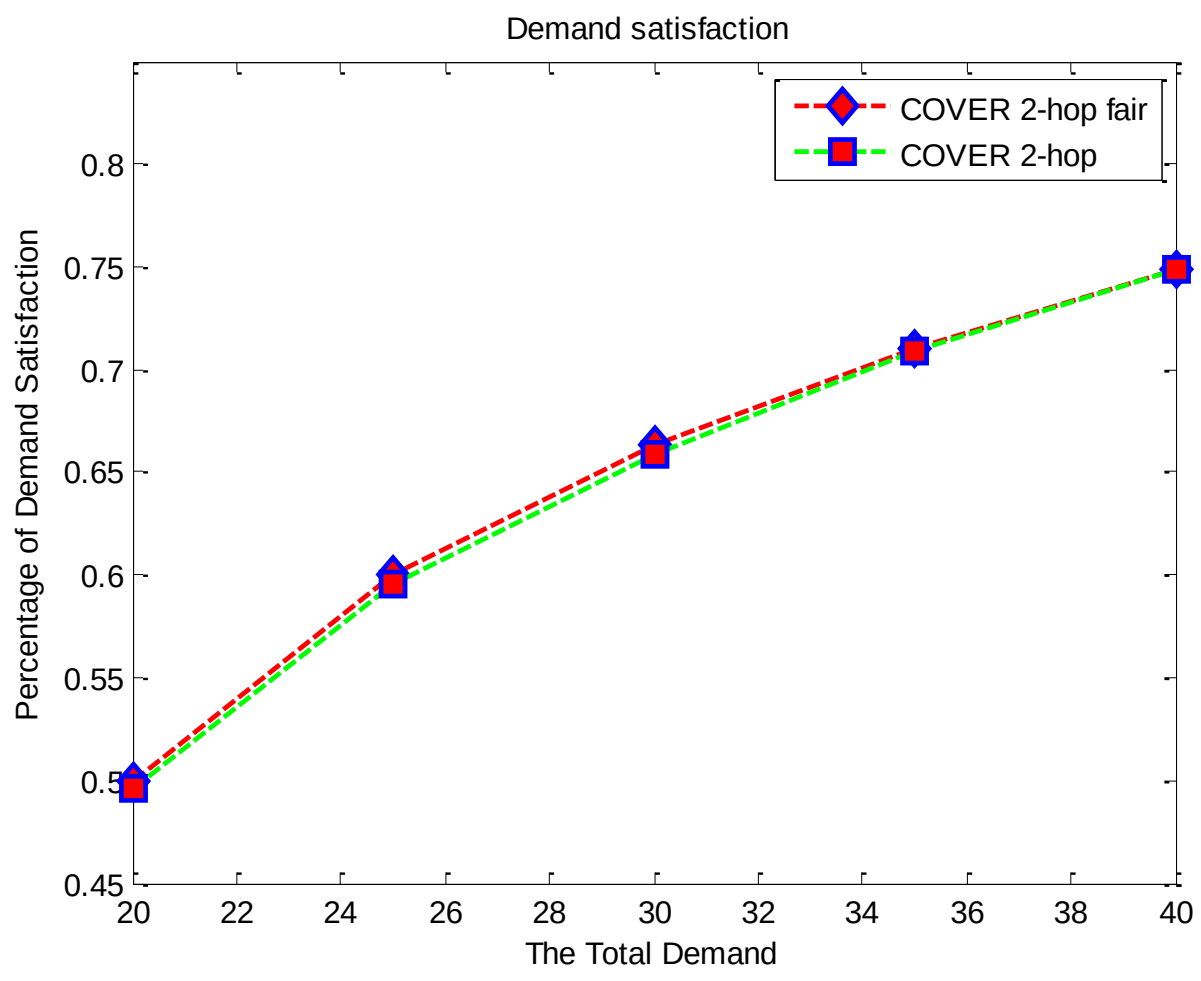

Figure 12: The level of demand satisfaction. The number of robots equals demand.

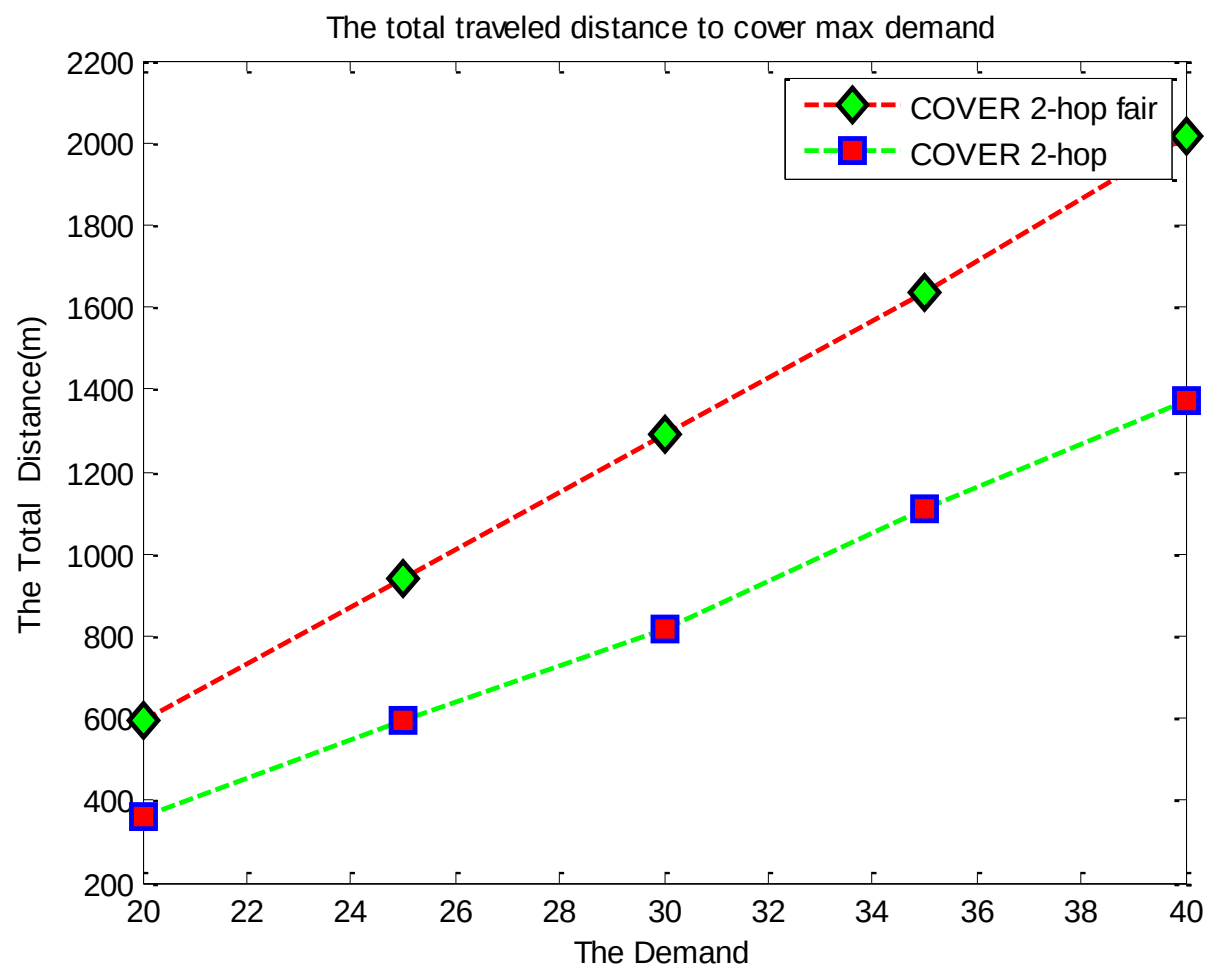

Figure 13: The total distance traveled to achieve the maximum possible demand satisfaction. The number of robots equals demand

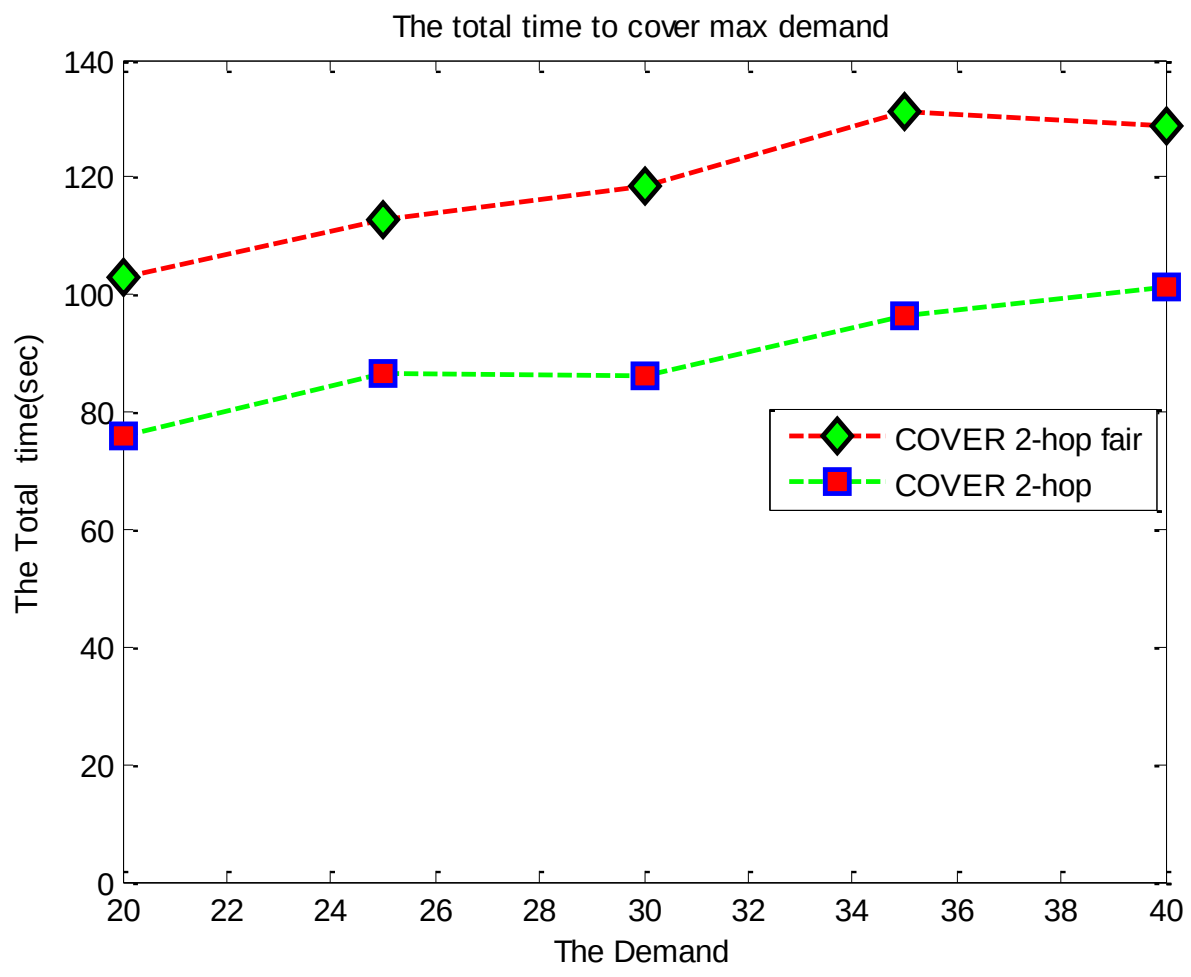

Figure 14: The total time needed to achieve the maximum possible demand satisfaction.The number of robots=demand-10

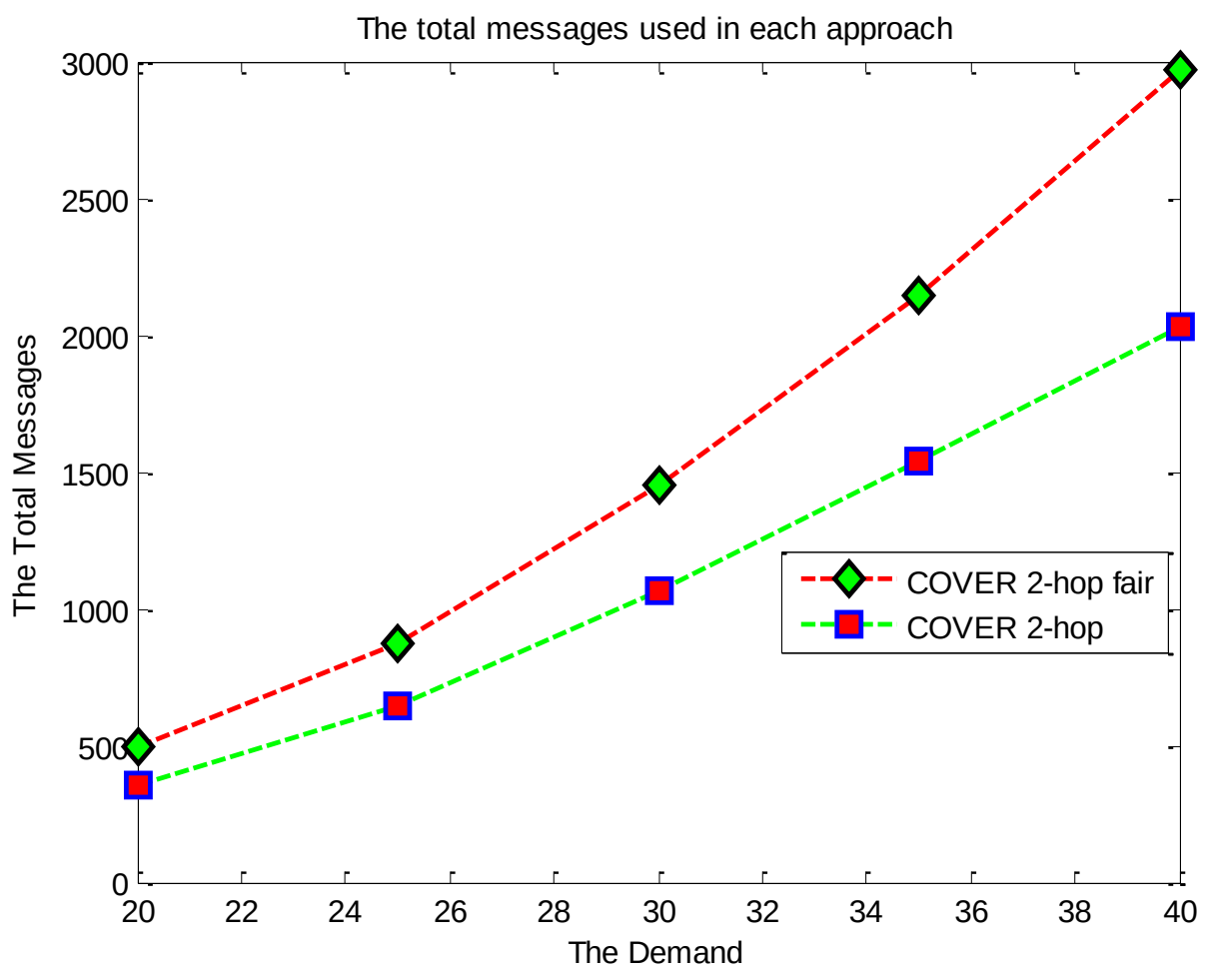

Figure 15: The total number of messages. The number of robots equals demand

Finally, the total number of messages increases by 20-50% in the new approach as in Fig 15. In the new approach, each landmark will start the cooperation earlier than in the previous approach. When a landmark achieved its minimum level of demand satisfaction it will start the cooperation while in the previous approach it will cooperate only when its demand is zero.

## V. CONCLUSION

In this paper, we have proposed a novel autonomous and cooperative distributive method for multirobot deployment using virtual force based on landmarks demand. Two-hop COVER approach was proposed as an improvement of COVER: it aims to shorten the time required, reduce the total distance and number of messages, and improve the level of demand satisfaction. Finally, since COVER and Two-hop COVER are not able to always reach 100% demand satisfaction, we proposed a Trace Fingerprint to do so in an efficient way. We considered the fairness in distributing robots among landmarks in case the total demand of the landmarks is greater than the available robots. As a future work, we need to implement the proposed algorithms on real robots. Moreover, robots failures in the middle of deployment and dealing with the case when the available robots are less than the demand are worth further investigations.

## ACKNOWLEDGMENT

The authors would like to acknowledge the support provided by the National Plan for Science, Technology and Innovation (MAARIFAH) - King Abdulaziz City for Science and Technology through the Science & Technology Unit at King Fahd University of Petroleum & Minerals (KFUPM), the Kingdom of Saudi Arabia, award project No. 11-ELE2147-4.